\documentstyle[multicol,pre,aps,epsf,psfig,epsfig]{revtex}     
\begin{document}  
\title
{
	Dynamical frustration in ANNNI model and annealing
}
\author
{
	Parongama Sen and Pratap Kumar Das
}
\address
{
	 Department of Physics, University of Calcutta,
	     92 Acharya Prafulla Chandra Road, Kolkata 700009, India.
}
\maketitle                  
\begin{abstract}
Zero temperature quench in the Axial Next Nearest Neighbour 
Ising  (ANNNI) model fails to bring it to its ground state for a certain range
of values of the frustration parameter $\kappa$, the
ratio of the next nearest neighbour antiferromagnetic 
interaction strength
to the nearest neighbour one. We  apply several annealing methods,
both classical and quantum,  and observe that the behaviour of the
residual energy and the order parameter depends on the value of $\kappa$
strongly. Classical or thermal annealing is found to be adequate for 
small values of $\kappa$.
 However, 
neither  classical nor quantum annealing is effective at values of $\kappa$
close to the fully frustrated point $\kappa=0.5$,  where the residual energy shows a 
very slow algebraic decay with the number of MCS.

\end{abstract}
\medskip



\begin{multicols}{2}

\section{Introduction}
Simulated annealing is usually applied to systems with frustration,
like spin glasses and optimisation problems, where the energy landscape is complex
with many
spurious minima. There are certain other systems, however, which have very simple 
energy landscape picture and ground states, but still the system fails to reach its 
ground state during a energy-lowering dynamical process. This situation corresponds
to ``dynamical frustration''. We have specifically considered 
the case of the axial next nearest neighbour (ANNNI) chain, where such a situation
is encountered.
In section II, we elaborate the notion of dynamic frustration with examples and in section
III, the dynamics in ANNNI model is discussed in detail. The results of application
of the classical and quantum annealing are discussed in sections IV and V.
Summary and some concluding comments are
given in the last section.

\section{Dynamic frustration in Ising models}

Quenching   dynamics in magnetic systems has been a topic of intense
research over the last  few decades. 
In quenching dynamics,  the system has a disordered initial configuration 
corresponding to
a high temperature. 
As the temperature is suddenly decreased
quite a few interesting phenomena take place like domain growth \cite{gunton,bray},
persistence \cite{satya1,derrida,stauffer,krap1} etc.

The Ising model maybe regarded as the simplest model
describing magnetic properties of many real systems and it shows a rich  
dynamical behaviour with respect to the above phenomena. The dynamics of 
Ising models has been
extensively studied in lattices of different dimensions as well as on
graphs and networks.
 
In dynamical studies, the system is allowed to evolve from the initial configuration following 
 a certain prescription and the commonly used dynamical rule at
   zero temperature is the Glauber dynamics, i.e., 
    a spin is chosen randomly
    and flipped if it makes the energy lesser, not flipped if energy
    increases and flipped with probability 1/2 if there is no energy change.

 \begin{center}
 \begin{figure}
 \noindent \includegraphics[clip,width= 4cm]{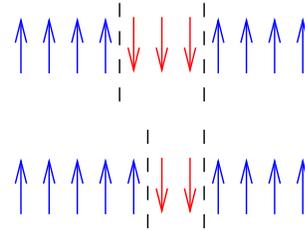}
 \caption{ Domain coarsening in one dimensional Ising model: the spin at
 the boundary of the left domain wall flips making the two domain walls
 move closer (Top: earlier picture, bottom: later picture).
 }
 \end{figure}
 \end{center}
   
 The zero temperature deterministic  
 dynamics in Ising models can be visualised as the motion of interfaces
  and the domains grow in size as the interfaces annihilate
  on approaching each other (Fig. 1). 
 In  one dimension, a zero temperature quench  of the  Ising model
 ultimately leads to the equilibrium
 configuration, i.e., all spins point up (or down).


  In two or higher dimensions, however, the system does not
  always reach equilibrium \cite{Krap_Redner,sundar,god,stein,jain,gleiser}. 
  Such a situation corresponds to dynamical frustration when the
  system gets frozen in a metastable state which does not correspond to the
  ground state.
  For example, in the
  two dimensional lattice (Fig. 2), the dynamics stops at a higher energy
  when the domain walls are straight and appear without any corner. 
  The 
  system thus acquires a ``striped phase'', where the number of stripes
  is an even number.
  In  dimensions higher than two, there maybe other 
  kind of frozen states in which the system gets locked.

  This kind of  freezing or blocking is also encountered in ferromagnetic 
Ising models on random graphs
   and small world  networks where there are  finite number of  random
   long range bonds. 
  In the random graph, any two spins are connected with a finite
  probability while 
  in the small world network,  random
  long range connections occur in addition to nearest neighbour links.
  In these cases, the domain walls may get pinned resulting in a frozen state \cite{sw1,sw2,sw3,sw4,sw5}.
  Recently, freezing has been observed on scale-free  networks also, where, 
  although the system is locked in an 
  excited state, the dynamics  continues indefinitely \cite{sw6}.
 \begin{center}
 \begin{figure}
 \noindent \includegraphics[clip,width= 4cm]{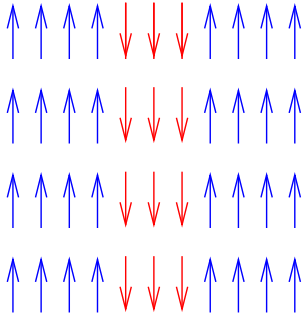}
 \caption{ Striped phase in two dimensional Ising model under zero temperature
 quenching dynamics - the domain walls are straight and no spin flips 
can occur.}
 \end{figure}
 \end{center}

  In the above examples of freezing in Ising models on finite
  dimensional lattices,  graphs and networks, a few things are to  be noted\\
  (a) The ground state is simple in all the cases\\
  (b) There is no frustration arising out of the interactions
  in the system.\\
In addition,  
   power law scalings with time (e.g., domain size $\sim t^{1/z}$) exist
  for the finite dimensional lattices \cite{comment} but in the case of  networks
or random graphs,
  one has an exponential relaxation behaviour consistent 
with the mean field nature of these systems \cite{silva,herrero,hong,lopes}.

\section{Dynamics in ANNNI chain}

We will now discuss the dynamics in the axial next nearest 
neighbour Ising (ANNNI) 
model \cite{selke} in which there is frustration but no
randomness or disorder. The ANNNI 
model in one dimension is described by the Hamiltonian 

\begin{equation}
H = -J_1\Sigma S_iS_{i+1} + J_2\Sigma S_iS_{i+2}.
\end{equation}

The ground state of this model is well-known:
it is ferromagnetic for $\kappa = J_2/J_1 < 0.5$;
antiphase  for $\kappa = J_2/J_1 > 0.5$
and
highly frustrated for $J_2/J_1=0.5.$
All configurations having domains of size $\ge 2$ are ground states at the
point $\kappa=0.5$, which we call the fully frustrated point.

The dynamics of the ANNNI chain has quite a few interesting behaviour \cite{redner,PS_SDG}.
When zero temperature quenching dynamics  dynamics is considered, 
 $\kappa = 1.0$ emerges as a dynamical transition point.
For $\kappa < 1$, there is no conventional
domain coarsening or persistence behaviour. Here,  
all domains of size 1 immediately vanish
but domains of size two are stable such that two domain walls cannot approach each other
and annihilate (Fig. 3).

Unlike in the Ising model in $d=2$, here the domain walls are not pinned but
can  move around keeping their number fixed.
The energy of the system is  constant as spin flips occur
with zero energy cost. The system thus wanders in a subspace of 
iso-energy metastable states forever.
As a result, the model also does not show a power-law decay in the
persistence probability. 
Interestingly, there is no special effect of the  $\kappa =0.5$ point (which
dictates the static behaviour) on the dynamics.

 \begin{center}
 \begin{figure}
 \noindent \includegraphics[clip,width= 3cm]{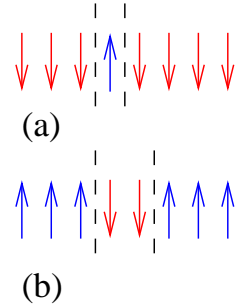}
 \caption{ Dynamics in ANNNI chain for $\kappa < 1$: All domains
 of size 1 are unstable and they immediately vanish (a). The
 system is left with domains of size $\ge 2$ . A domain of size two,
 as shown in (b), is energetically stable as the spins within the domain
 satisfy the antiferromagnetic ordering with their second neighbours and  
 energy contribution from first neighbours is zero.
 As a result,
 two domain walls have to maintain a minimum distance and 
 cannot approach each other and annihilate. 
 }
 \end{figure}
 \end{center}

As the domain walls continue moving in the system, the number of
spin flips at any time becomes a constant in time. This constant is independent 
of the value of $\kappa$.
That is expected as this 
quantity is proportional to the number of domain walls. The average number of
domain walls remaining in the system (per spin) turns out to be close to 0.28\cite{PS_SDG}.

\begin{center}
 \begin{figure}
  \noindent \includegraphics[clip,width= 4cm, angle=270]{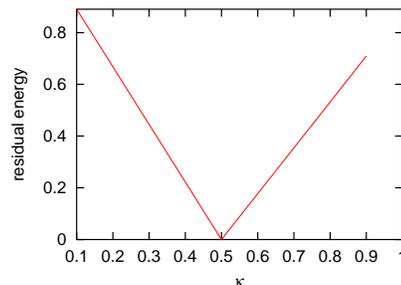}
   \caption{ 
     Constant residual energy of the ANNNI model with zero temperature
     dynamics for $\kappa < 1$.}
      \end{figure}
       \end{center}

The  residual energy $E_r$, defined as the excess energy over the 
ground state energy shows an interesting behaviour with $\kappa$ (Fig. 4).
At small values of $\kappa$ the large number of domain 
walls makes the residual energy large. As $\kappa$ is increased,
$E_r$ becomes lesser and at $\kappa=0.5$ it is zero.
This is not surprising however; the configurations with domain 
sizes $\ge2$ are
nothing but the degenerate ground states of the $\kappa=0.5$ point.
The residual energy decreases as $\kappa = 0.5$ is approached from both sides.

For the sake of completeness, it should be mentioned that
the ANNNI model shows conventional relaxation behaviour 
for $\kappa > 1$ although with a dynamic exponent 
	different from that of the nearest neighbour Ising model.

Since the conventional dynamics does not bring the system 
to the ground state, one has to employ some other method
to do this. 
Simulated annealing for the Ising model on random graphs has been
attempted to ``melt'' the system   with  success \cite{sw1}.
For small world networks, freezing can be got rid off by 
letting more number of edges in the system also \cite{pkd}.
We discuss in the next two section the result of 
applying different annealing schedules to the ANNNI model
for various values of $\kappa < 1$.

\section{Classical Annealing (CA)}

        We have adopted two different schemes for applying classical or thermal
annealing.

\medskip
{\it {Scheme A}}
\medskip

 This is the conventional  scheme where one  starts
 with a finite temperature $T=T_0$
and slowly reduces it according to a linear schedule, such that,
at the $t$-th iteration step,

\begin{equation}
T= T_0(1-t/\tau),
\end{equation}
where $\tau$ is the total number of Monte Carlo steps (MCS).
The final temperature (at $t=\tau$) is zero for any starting value of $T$.

Since for all non-zero temperature, the ANNNI model is in a paramagnetic
state, one may  start with a random initial configuration corresponding to
$T_0$. 

We have calculated the residual energy
  and  the order parameter as functions of $\tau$; the former
  is expected to approach zero  and the latter
  should increase towards unity with larger values of $\tau$. 



In Fig. 5, we show the behaviour of $E_r$ against $\tau$ for different $\kappa$
values. These  simulations have been done for a system of 100 spins
keeping  $T_0=10$. 
 The number of  configurations $n$ over which 
averaging has been done decreases with  $\tau$; starting with $n=1000$ for
smaller values of $\tau$, 
it is decreased to $n=100$ for  very large $\tau$ values. 
While for $\kappa =0.2$, we find a stretched exponential decay,
the nature of the curves changes to a power law decay 
for higher values of $\kappa$. Close to
0.5  it has a very slow decay. 
Corresponding to a   power law fit $E_r \sim \tau^{-\alpha}$,  
$\alpha$ is very small here, e.g., for $\kappa=0.4, ~~\alpha = 0.08\pm 0.01,$
and for $\kappa = 0.6, ~~\alpha = 0.03\pm 0.01 .$  
The slowing down of the decay pattern of $E_r$ seems to depend on
the closeness to $\kappa =0.5$ as well as on the nature of the ground state.
 For the various values of $\kappa$ for which the annealing scheme has been
employed, the slowest decay 
 is observed at $\kappa=0.6$, when
the system is close to $\kappa=0.5$ and the ground state is also antiphase. 
The power law remains valid for  $\kappa > 0.5$ with  an increasing 
value of $\alpha$.
\begin{center}
\begin{figure}

\includegraphics[clip,width= 6cm, angle=270]{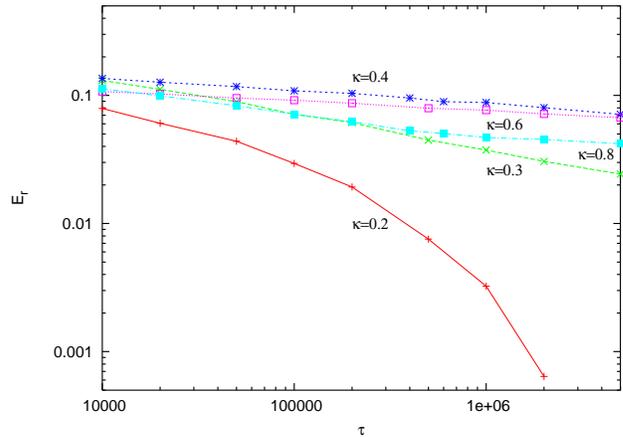}
\caption{Residual Energy vs. MCS $\tau$ in the ANNNI model 
for different values of $\kappa < 1.0$. The decay is slow close to 
$\kappa=0.5$. $T_{0}=10$ here.}
\end{figure}
\end{center}


We have also checked the efficiency of this scheme for different values of 
$T_0$. Note that the decrease in $T$ is made with a slope equal to $T_0/\tau$ 
(eq. 2),
so we compare the results for different $T_0$'s by plotting
$E_r$ against $\tau/T_0$. For $\kappa=0.4$ or $0.6$, $E_r$ is independent
of $T_0$ (as long as $T_0$ is not very small compared to 1)
while for $\kappa=0.2$, lowering  $T_0$ makes the decay  faster at
large $\tau$. However, when  $T_0$ is made smaller than unity 
the decay does not become  faster any more. This plot (Fig. 6) also  shows 
that for $\kappa$ close to 0.5,
the power law behaviour is actually valid over a large range of $\tau$
which is not so apparent from Fig. 5.
\begin{center}
\begin{figure}
\includegraphics[clip,width= 6cm, angle=270]{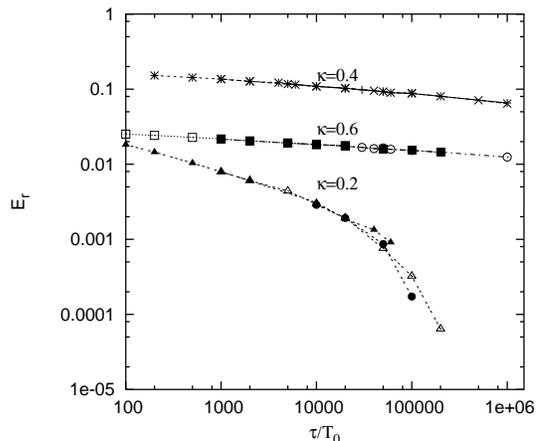}
\caption{The residual energy vs the number of MCS ($\tau$) for
different starting values of $T_0$. The scaled data shows a good
collapse for $T_0\geq 1$ for $\kappa=0.4,0.6$. For
$\kappa=0.2$, the decay is fastest for $T_0=1$ (filled circles). The data for $\kappa = 0.6$ and $0.2$ have been
shifted for clarity.}
\end{figure}
\end{center}

Thus $E_r$ depends strongly on $\kappa$: not only does the functional form 
change from a stretched exponential to power law, there is also a non-universal
exponent $\alpha$ which depends on the value of $\kappa$.
 The role of $\kappa=0.5$ is
felt clearly as the annealing is least effective close to this point.

While estimating the order parameter (OP), it should be noted that 
 for $\kappa < 0.5$, the order parameter is just the
magnetisation while for $\kappa > 0.5$, it is the average of
the four sublattice magnetisations defined as 
\begin{equation}
m_\alpha = \Sigma_{j=0}^{L/4-1} S_{\alpha + 4j}; ~~~\alpha =1,2,3,4
\end{equation}    
as in \cite{PS_SDG}.
In Fig. 7, 
we plot the behaviour of the OP with $\tau$ for different values of $\kappa$.
The behaviour of the order parameter is also $\kappa$
dependent. The variations with $\tau$ are not smooth: the reason is that
decreasing energy is not necessarily  equivalent to increasing  the 
order parameter. 
However  it seems to have  a rough power law increase for all values of $\kappa$ (Fig. 7).
As expected, the growth of the order parameter is slowest at $\kappa=0.6$.
\begin{center}
\begin{figure}
\includegraphics[clip,width= 6cm, angle=270]{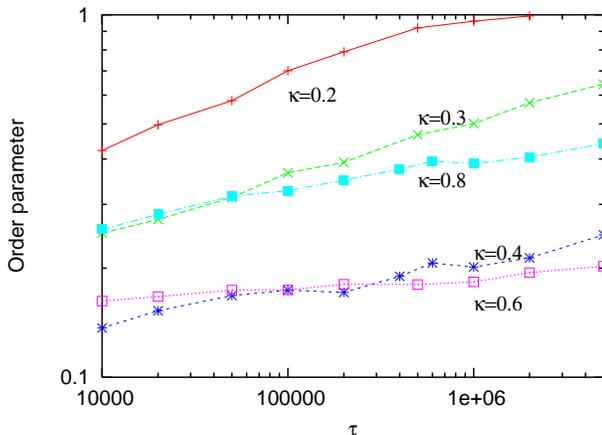}
\caption{Order parameter  vs. MCS $\tau$ in the ANNNI model 
for different values of $\kappa < 1.0$. The growth is slow close to $\kappa=0.5$.}
\end{figure}
\end{center}

Thus we find that this annealing schedule is not very effective near $\kappa=0.5$ 
but works well for small values of $\kappa$. It is not possible to
detect  whether there is a value of $\kappa$ for which the behaviour of the
residual energy changes from stretched exponential to power
law from the present numerical study.

\medskip
{\it Scheme B}

\medskip
In this scheme, we first let the system evolve from a random configuration
(corresponding to a high temperature)
using the zero temperature dynamics and then, after a few steps,
apply scheme A. The difference here is, we let the system reach one of the
metastable states under zero temperature dynamics, heat it to a finite 
temperature
$T_0$ which is then gradually decreased.

	In the ANNNI chain, for the first 100 iterations, the 
	temperature is kept zero such that when re-heated,
the system is in one of the metastable iso-energy states.
We find that this scheme 
accelerates the decay of the residual energy remarkably
for small values of $\kappa$ (e.g., $\kappa=0.2$) at large $\tau$. 
However, for higher values, e.g., $\kappa =0.3,0.4, 0.6$, for which 
scheme A gave a power law decay of $E_r$, the results are identical to 
that of scheme A. 
Thus near the $\kappa=0.5$ point,  this scheme is also seen to fail to bring the system
to its static ground state.

This scheme is highly appropriate for cases
where the system has a fractional probability to end up
in a metastable or frozen state not
corresponding to the ground state.
 Here it is  not possible 
to predict  whether a random initial configuration will reach
the ground state under zero temperature dynamics or not.
A good example is the two dimensional
Ising model as it reaches a frozen state in about 30\% cases.
It is useless to apply scheme A here because 70\% of 
the cases do not require any annealing at all.
Therefore in order to see whether CA is useful,  
it should be better to apply   scheme A to an
initial configuration of frozen state 
which can be  assumed to have evolved from a perfectly 
random initial state with zero temperature Glauber dynamics.
Thus effectively it has undergone a period of cooling at zero temperature
and when scheme A is now applied to it, it is equivalent to scheme B.
\begin{figure}
\includegraphics[clip,width= 6cm, angle=270]{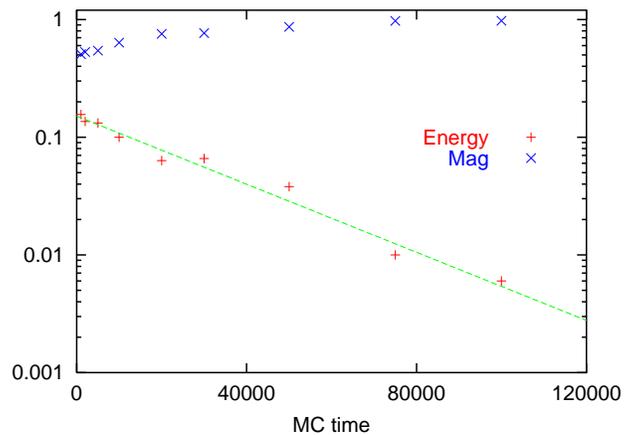}
\caption{Order parameter (magnetisation) and residual energy versus MC time $\tau$  in the 2-$d$ Ising  model
under scheme B. The value of $T_0$ is 1 here. The dashed line has slope equal
to $3.33\times 10^{-5}$
in the log-linear plot.}
\end{figure}

We have performed some simulation on a square lattice of size $L=40$, where the number of stripes  $s$ is
equal to 2 or 4.
 In each  case, we find that the behaviour of both the 
residual energy and the magnetisation is exponential which means that the scheme
works very well in this case. In Fig. 8, we show the variation of the 
residual energy and magnetisation for $s=2$.
The exponential relaxation is easily 
understood, as the thermal perturbation breaks the 
structure of the domain walls and the system is then again free to
evolve dynamically. The results are identical for $s=2$ and $4$ at large $\tau$. It may be mentioned here that for a value
of the stripe number $s$ comparable to $L$, the situation is very similar
to the ANNNI model (as stripe sizes have to be $\ge 2$)
and then the exponential behaviour may no longer be present. However, the 
probability of a large value of $s$ is small and so, we have not considered
values of $s > 4$.

\section{Quantum Annealing (QA)}

Although the classical annealing methods work quite well for the
ANNNI model for small $\kappa$, we find that close to the $\kappa=0.5$
point, it leads to very slow relaxation.
   In several situations, quantum annealing is far more efficient
   in decreasing the energy of the system \cite{new2,sontoro} and we therefore
   apply this method in the ANNNI model.
Here instead of thermal fluctuation,  quantum fluctuation is considered to 
induce tunnelling to enable the  system
reach the ground state for $\kappa < 1$.
  
  The Hamiltonian for the quantum ANNNI chain is :
  \begin{equation} 
   $$H=-J_1\Sigma S_{i}S_{i+1} + J_2\Sigma S_ iS_{i+2}
   -\Gamma \Sigma S_ i$$
    \end{equation}

   This   can be mapped to a 2-dimensional classical model \cite{suzuki,CDS}
 using the Suzuki-Trotter formula:
    
    \begin{equation} 
     H=-J_1\Sigma S_{\alpha, i}S_{\alpha,i+1} + J_2\Sigma S_{\alpha, i}S_{\alpha ,i+2} -J_p\Sigma S_{\alpha, i}S_{\alpha+1,i}
     \end{equation} 
      where
      
     $$ J_p = - {PT/2}\ln (\tanh (\Gamma/PT))$$  
and $\alpha$ denotes the $\alpha th$ row in the Trotter direction.
 Subsequently, one can use a linear schedule for the transverse field
as in \cite{sontoro}  and find out the results for $E_r$ and the order parameter.
 However, it is not possible to make $\Gamma$ equal to zero in the 
 last step as that would make $J_p$  infinite.
 
 We would first show some curious features of the results
 on applying this method to the ANNNI model and then try to justify the 
 results.

 The Suzuki-Trotter mapping is  exact for $P \to \infty$ 
 but it can be a good approximation if $PT \ge 1$.
 One needs to find out an optimum value of $PT$ for which
 $E_r$ does not change with $P$.
 We therefore fix $PT$ and find out $E_r$ for
 different values of $P$ following \cite{sontoro}. We first fix $PT=1$. 
 For small $\kappa$, e.g., for $\kappa=0.1$,
 the scheme indeed makes $E_r$ go down with $\tau$ quite
 efficiently. However, the value of $E_r$ for the same $\tau$ and
 different values of $P$ shows that $E_r$ actually increases with $P$.
 Thus results for any finite $P$ may not be reliable. 
 Even increasing $PT$ to 2, we find that this behaviour persists.
 The reason for this maybe that the Suzuki-Trotter mapping works with 
 a non-zero temperature 
for which the system is disordered and is always at a high energy state
	compared to the perfectly ordered state and therefore
	$E_r$ does not go to zero
	for large $\tau$ and $P$. These results are shown in Fig. 9.
\begin{center}
\begin{figure}
\includegraphics[clip,width= 6cm, angle=270]{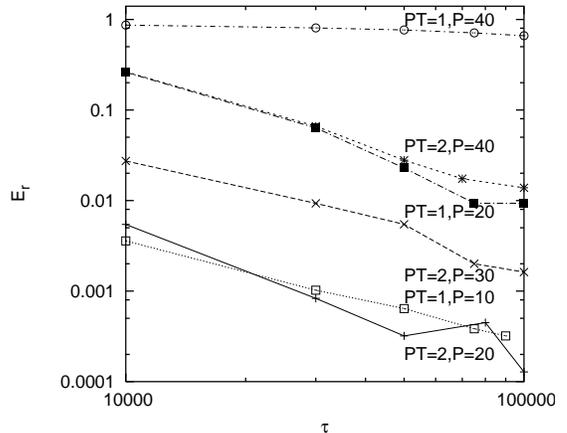}
\caption{The quantum annealing scheme does not appear to work well
for the ANNNI chain. For example at $\kappa=0.1$,
an optimum value of $PT$ is difficult to find out as $E_r$ increases with $P$ 
 where $PT$ is fixed at $1$ or $2$ at any value of $\tau$. }

\end{figure}
\end{center}

\begin{center}
\begin{figure}
\includegraphics[clip,width= 6cm, angle=270]{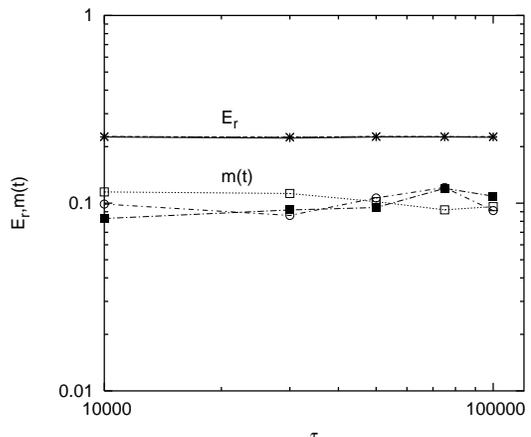}
\caption{The variation of residual energy and order parameter for $\kappa=0.4$
against $\tau$ are shown for $3$ different values of $P$. $E_r$ actually remains
constant with $\tau$ and $P$ while the $OP$ (m(t)) seems to fluctuate around a
constant value.}
\end{figure}
\end{center}

In case of a value of $\kappa $ close to 0.5, e.g., $\kappa=0.4$,
we find that $E_r$ remains virtually a constant for all $P$ values when $PT=1$
which apparently implies that $PT=1$ is a good optimum value. 
However, $E_r$ actually remains a constant for all $\tau$ values as well 
showing that it does not relax at all.
Thus here too the quantum annealing method will  not work well.
The reason is again because a non-zero temperature of the system has been used.
However, the manifestation of this non-zero temperature is different
for small and higher values of $\kappa$.

On hindsight, it may appear that quantum annealing is a redundant
exercise in this case. However, it is interesting to find out 
how the redundancy makes itself known for different values of $\kappa$ in 
different ways.

\section{Summary and conclusions}

In summary, we have shown that  in systems with dynamic frustration,
simulated annealing can be applied which gives results according to the
nature of the system. For the ANNNI model, which has a competing
interaction leading to frustration, (but well defined ground states
with trivial degeneracy for $\kappa \ne 0.5$)
 classical annealing seems to work well 
for values of $\kappa$ close  to $\kappa=0$. 

We have applied a different scheme of thermal annealing 
where the system is heated after an initial period of cooling. 
The results remain same in case of the ANNNI model when the frustration
parameter is appreciable. But this 
method is useful in case of some other models where the
conventional scheme is not very handy, e.g., the two dimensional
Ising model. The better effectiveness  of the annealing scheme 
for the two dimensional model in comparison to
	the ANNNI model may be attributed to the
fact that there is no frustration in the former.
 The frustration present in the
ANNNI model especially near $\kappa=0.5$ makes it unresponsive to the thermal annealing while
the unfrustrated two dimensional Ising model responds fast to the
thermal fluctuation. 

In case of the ANNNI model, it is also interesting to note that while 
in the dynamical studies it was shown that the $\kappa=0.5$ point
hardly has any role to play, and dynamical 
quantities like persistent probability, number of spins flipped
at any time etc. were $\kappa$ independent, things become 
strongly $\kappa$ dependent under any annealing schedule with non-universal
exponents governing the power law  decays. The effect of $\kappa=0.5$ 
is also felt as the annealing fails to make any  impact close to it. 

The application of a quantum annealing method with non-zero temperature 
also fails to "melt"
the system to its ground states and its effect is differently manifested
for small and large values of $\kappa$s.

In fact we find that much is left to be done for a successful annealing 
programme near $\kappa=0.5$, e.g., one may attempt a zero-temperature
quantum annealing schedule. 

Even for the classical annealing case, the change in behaviour of
the residual energy as $\kappa$ is varied requires to be studied 
more intricately and possibly for larger syatem sizes. Our
present study may act as a guideline for such future research work.

Lastly, we note that the variations of the residual energy does
not follow a Huse-Fisher \cite{huse} type scaling in any parameter range
for the ANNNI model.

Acknowledgements: We are grateful to Arnab Das and Subinay Dasgupta for very valuable discussions.
P. K. Das acknowledges support from CSIR grants no.
9/28(608)/2003-EMR-I.

\end{multicols}

\begin{thebibliography}{references:}
\bibitem{gunton} J. D. Gunton, M. San Miguel and P. S. Sahni, {\it{Phase
Transitions and critical phenomena}}, Vol 8, eds. C. Domb and J. L. Lebowitz (Ac
\bibitem{bray} A. J. Bray, {Adv. Phys. {\bf{43}} 357 (1994) and the references
therein}.
\bibitem{satya1} For a review, see S. N. Majumdar, {Curr. Sci.
{\bf{77}} 370 (1999)}.
\bibitem{derrida} B. Derrida, A. J.Bray and C. Godreche, {J.Phys. A {\bf{27}}
L357 (1994)}
\bibitem{stauffer} D. Stauffer, {J. Phys. A {\bf{27}} 5029 (1994)}.
\bibitem{krap1} P. L. Krapivsky, E.Ben-Naim and S. Redner,
{Phys. Rev.E {\bf{50}} 2474 (1994)}.
\bibitem{Krap_Redner} V. Spirin, P. L. Krapivsky and S. Redner,
{Phys. Rev.E {\bf{63}} 036118 (2001)};
Phys. Rev.E {\bf{65}} 016119 (2002).
\bibitem{sundar} P. Sundaramurthy and D. L. Stein, cond-mat/0411286.
\bibitem{god} C. Godreche and J. M. Luck, cond-mat/0412077.
\bibitem{stein} C. M. Newman and D. L. Stein, {Phys. Rev. Lett. {\bf{82}}
3944 (1999)}.
\bibitem{jain} S. Jain, {Phys. Rev.E {\bf{60}} R2445 (1999)}.
\bibitem{gleiser} P. M. Gleiser, F. A. Tamarit, S. A. Cannas and M. A. Montemurro,
 {Phys. Rev. B {\bf{68}} 134401 (2003)}.

\bibitem{sw1} P. Svenson, Phys. Rev. E {\bf {64}} 036122 (2001).
\bibitem{sw2} O. Haggstrom, {Physica A {\bf{310}} 275  (2002)}.
\bibitem{sw3} D. Boyer and O. Miramontes, {Phys. Rev. E {\bf {67}}
 R035102 (2003)}.
 \bibitem{sw4} P. Svenson and D. A. Johnson, {Phys. Rev. E {\bf {65}}
  036105 (2002)}.
  \bibitem{sw5} J. Y. Zhu and H. Zhu, {Phys. Rev. E {\bf {67}} 026125
   (2003)}.
   \bibitem{sw6} D. Jeong,  M. Y. Choi and H. Park, cond-mat/0501099.
   \bibitem{comment} The persistence probability in case of four dimensional
   lattice, however, does not show a power law decay due to blocking 
   (D. Stauffer, Int. J. Mod. Phys. C {\bf 8} 361 (1997)).
   \bibitem{silva} N. R. da Silva and J. M. Silva, {Phys. Lett. A {\bf{135}}
   373 (1989)}.
\bibitem{herrero} C. P. Herrero, {Phys. Rev. E {\bf {65}} 066110 (2002)}.
\bibitem{hong} H. Hong, B. J. Kim and M. Y Choi, {Phys. Rev. E {\bf{66}} 011107
(2002)}.
\bibitem{lopes} J. V. Lopes etal, cond-mat/0402138.
   \bibitem{selke} W. Selke, {Phys. Rep. {\bf {170}} 213 (1988)}.
    \bibitem{redner} S. Redner and  P. L. Krapivsky, J.Phys. A {\bf 31} 9229 (1998)
\bibitem{PS_SDG}  P.Sen and S.Dasgupta, J. Phys. A {\bf 37} 11949 (2004) 
\bibitem{pkd} P. K. Das and P. Sen, cond-mat/0503138. 
\bibitem{new2} T. Kadawoki and H. Nishimori, {Phys. Rev. E {\bf 58} 5355 (1998)}. 
\bibitem{sontoro} G. E. Santoro, R. Martonak, E. Tosatti and R. Car, Science {\bf 295} 2427 (2002).
\bibitem{suzuki} M. Suzuki, Prog. Theor. Phys. {\bf 56} 1454 (1976). 
\bibitem{CDS} B. K. Chakrabarti, A. Dutta and P. Sen, Quantum Ising Phases  and Transitions in Transverse Ising models, Lecture
Notes in Physics {\bf M41},  Springer-Verlag, 1996. 
\bibitem{huse} D. A. Huse and D. S. Fisher, Phys. Rev. Lett. {\bf 57} 2203 (1986).




  
                                                        
             


\end{thebibliography}
\end{document}